\documentclass[prc,superscriptaddress,showpacs,twocolumn,floatfix]{revtex4-1}
\usepackage{graphicx}
\usepackage{dcolumn}
\usepackage{bm}
\usepackage{amsmath} %for text inside the equation, 
\usepackage[utf8]{inputenc}
\usepackage[top=1cm,left=1cm,right=1cm,bottom=1cm]{geometry}% margins

\usepackage{url}											% URLs
\usepackage[usenames,dvipsnames]{xcolor}					% color									% columns env.
\usepackage{paralist}										% compact lists

\usepackage{changepage}
\usepackage[breaklinks=true]{hyperref}
\usepackage{mathtools}
\usepackage{braket}
\usepackage{amsmath}
\usepackage{cancel}

\begin{document}
	\title{Model studies on the COVID-19 pandemic in Sweden}
	\author{Chong Qi}\email{chongq@kth.se}
	\affiliation{Royal Institute of Technology at Albanova,
		SE-10691 Stockholm, Sweden }
	\author{Daniel Karlsson, Karl Sallmen, Ramon Wyss}
	\affiliation{Royal Institute of Technology at Albanova,
		SE-10691 Stockholm, Sweden }
	\date{\today}
	\begin{abstract}
		We study the increases of infections and deaths in Sweden caused by COVID-19 with several different models: Firstly an analytical susceptible-infected (SI) model and the standard susceptible-infected-recovered (SIR) model. Then within the SIR framework we study the  
		susceptible-infected-deceased (SID) correlations. All models reproduce well the number of infected cases and give similar predictions. What causes us deep concern is the large number of deaths projected by the SI and SID models. Our analysis shows that, irrespective of the possible uncertainty of our model prediction, the next few days can be critical for determining the future evolution of the death cases.  
		
	\end{abstract}

	\maketitle

\section{Introduction}
The fast spread of COVID-19 (nearly 1M infected cases till April 2nd, 2020) has caused wide concern. Within the basic research community, quite a few mathematical and physical models have been proposed \cite{Ising,Dmodel,Cal,Imp} to study the evolution of the infected cases, aiming to make reliable predictions and to help the governments to make proper strategic preparedness and response plans. We deem it is of special importance to study the COVID-19 spreading in Sweden where, unlike other countries, the government is  taking a rather relaxed strategy with no massive testing on suspected individuals and no strict lockdown in her most affected regions.

We start by introducing the SIR (susceptible-infected-recovered) model which is widely used for virus spreading predictions \cite{SIR,SIR2}.  It
consists of a system of three time-dependent variables:
\begin{itemize}
  \item Infected cases (number of total infected individuals at given time), $I(t)$.
  \item Susceptible cases (number of individuals susceptible of contracting the infection), $S(t)$.
  \item Recovered cases (cumulative number of recovered individuals), $R(t)$.
\end {itemize}
One has in total  $N=  S(t)+I(t)+ R(t)$. Above quantities satisfy the following non-linear differential
equations
\begin{eqnarray}
  \frac{dS}{dt} & = & - \lambda S I \nonumber\\
  \frac{dI}{dt} & = &  \lambda S I -\beta I \nonumber\\
  \frac{dR}{dt} & = & \beta I 
 \end{eqnarray}
where $\lambda$ is the transmission rate, $\beta$ the
recovery rate. They can be parametrized using known infection data.

It should be straightforward to extend the models to include deceased cases in the form
\begin{eqnarray}
\frac{dD}{dt} & = & \nu I 
\end{eqnarray}
and the exposed cases (individuals who are already infected but asymptomatic)
\begin{eqnarray}
\frac{dE}{dt} & = & \lambda S I -\sigma E
\end{eqnarray}

\section{The SI model}
If the recovery rate is very low during the pandemic time interval (as it is indeed the case for COVID-19 upto now), we can well approximate the infected cases by
\begin{equation}
  \frac{dI}{dt}  \approx   \lambda \left[(N-I(t)\right] I(t)\\
\end{equation}
with initial value $I(0)=1$.
For such SI model, the solution can be derived analytically,
\begin{equation}
  I(t) = \frac{ N }  {  1+(N-1)  {\rm e}^{-\lambda N t}  }
\end{equation}
This function is known as Woods-Saxon form in nuclear physics and is widely used describing nuclear potential and matter distribution. We can write it in a more general form by introducing one more parameter as

For such SI model, the solution can be derived analytically,
\begin{equation}
I(t) = \frac{ N }  {  1+{\rm e}^{{(t_0- t)}/{d}}  }
\end{equation}
where $d$ describes the diffuseness.

\begin{figure}
	\begin{center}
		\includegraphics[width= 9.5cm]{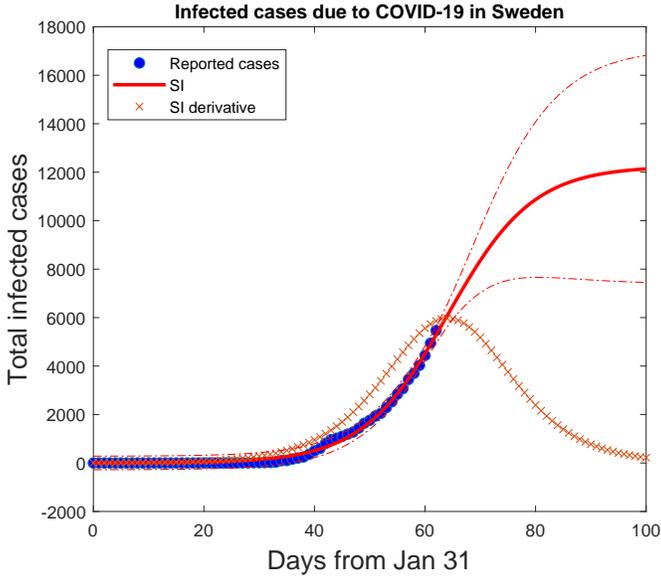}\label{total}
		\caption{SI description of total infected cases in Sweden as of April, 02. The total projected case is $N=12250$. Dash-dotted line shows the distribution within $95\%$ confidence level. The cross symbols shows the derivative of the SI model in random scale. The peak of the derivative (i.e., predicted day of maximum increase in reported cases) is on April 03. Data from \cite{Data}.
		}
	\end{center}
\end{figure}

We apply above formula to study the reported cases in Sweden as a function of time in Fig. \ref{total}.

\begin{figure}
	\begin{center}
		\includegraphics[width= 9.5cm]{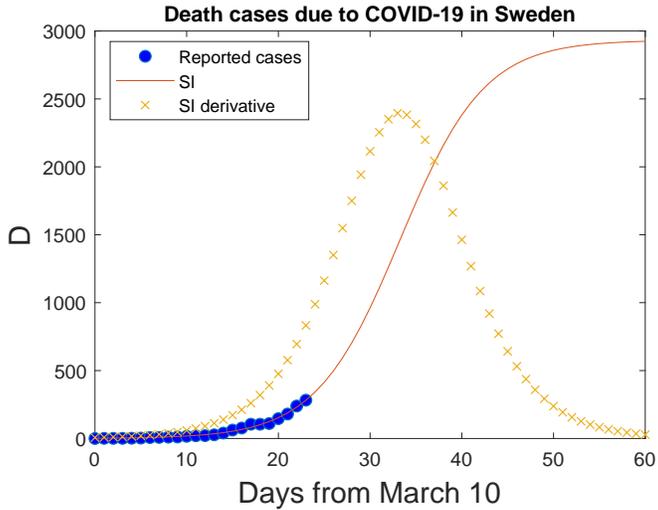}
		\caption{SI description of total death cases in Sweden as of April, 02. The total projected case is $N=2933$ with however, very large uncertainty.  The cross symbols shows the derivative of the SI model in random scale. The peak of the derivative is on April 12. Data from \cite{Data}.
		}\label{Death}
	\end{center}
\end{figure}

\begin{figure}
	\begin{center}
		\includegraphics[width= 9.5cm]{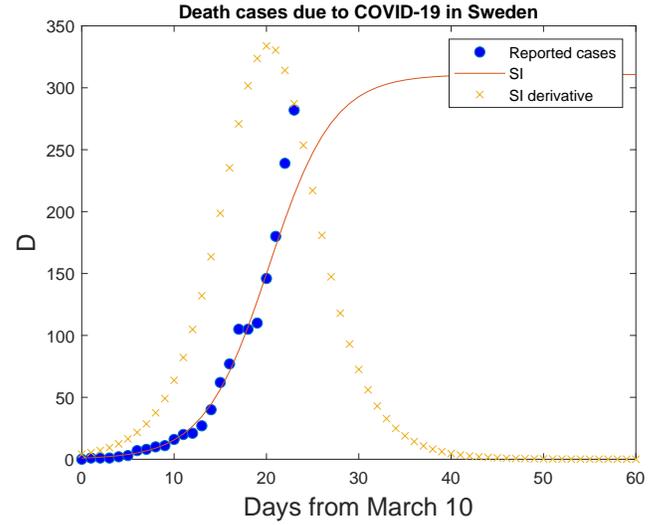}
		\caption{Same as Fig. \ref{Death} but the prediction curve is determined with data until March 31.\cite{Data}.
		}\label{Deathold}
	\end{center}
\end{figure}

\begin{figure}
	\begin{center}
		\includegraphics[width= 9.5cm]{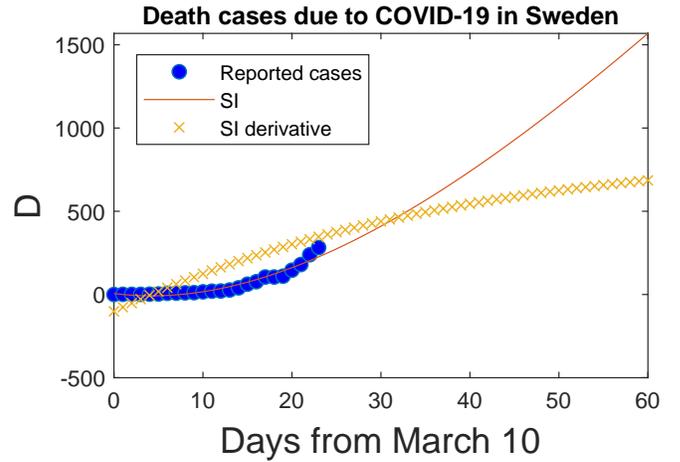}
		\caption{Same as Fig. \ref{Death} but with the prediction curve Eq.(\ref{d3}).
		}\label{Deathv2}
	\end{center}
\end{figure}

We firstly assume that the death cases follow the same Woods-Saxon form
\begin{equation} \label{}
 D(t) = \frac{ D_0 }  {  1+{\rm e}^{{(t_1- t)}/{d_d}}  }\end{equation}
where the three parameters can be fitted separately to reported data by taking into account the time interval between infection and death dates. The result is given in Fig. \ref{Death}. We should emphasize that the uncertainty in the model is very large due to the limited data available. We were very optimistic when we first derived the curve from data back by two days which was very different with modest increases, as can be seen in Fig. \ref{Deathold}.

The above Woods-Saxon function seems to agree rather well with the data on reported COVID-19 death cases from China where the pandemic period may be expected to be over. It may, however, not expected to work well within the late stage of the pandemic period when the total infected cases show a saturation behavior. Therefore we insert Eq. (6) to Eq. (3) to see if we can get a better estimation on $D$. The integral of $I(t)$ reads
\begin{equation}
\int I(t)=N\left(d \ln \left(\mathrm{e}^{\frac{t_0-t}{d}}+1\right)+t\right)
\end{equation}
Therefore, we can propose a second form for the evolution of $D$
\begin{equation}\label{d3}
D(t)=a \ln \left(\mathrm{e}^{\frac{t_0-t}{d}}+1\right)+bt+c
\end{equation}
The result is given  in Fig. \ref{Deathv2} where a modest increase is predicted. However, again, the data from the last two days show significant deviation from the predicted curve. New data from the next few days can be critical in pining down the uncertainty in the predicted behavior.  

\section{The SIR model}
\begin{figure*}
	\begin{center}
		\includegraphics[width= 19.5cm]{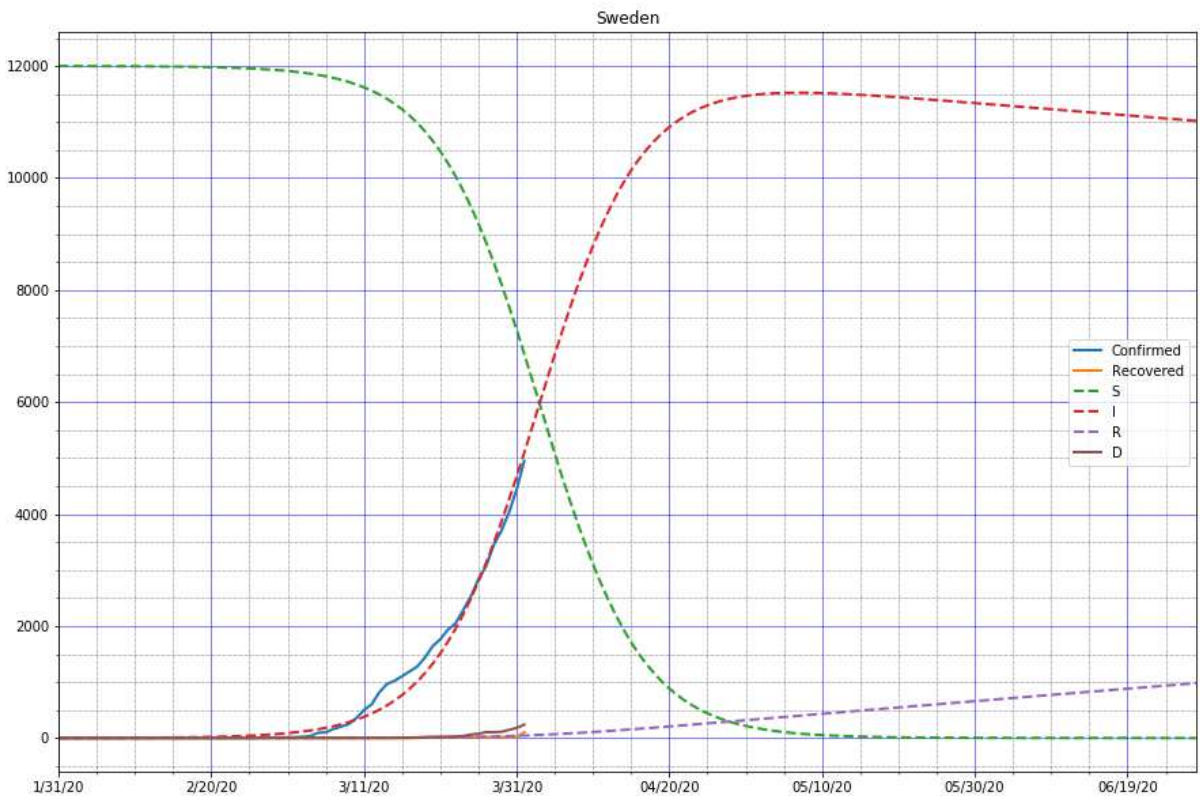}
		\caption{SIR model description of COVID-19 infected and recovery cases in Sweden as of April, 01. We take $N=12000$.
		}\label{sir}
	\end{center}
\end{figure*}

We now include the recovery cases in above SI model. There is no analytical solution but the evolution of the SIR quantities can be done numerically. The result as of April 01 is given in Fig. \ref{sir}

\section{The SID model}
\begin{figure*}
	\begin{center}
		\includegraphics[width= 19.5cm]{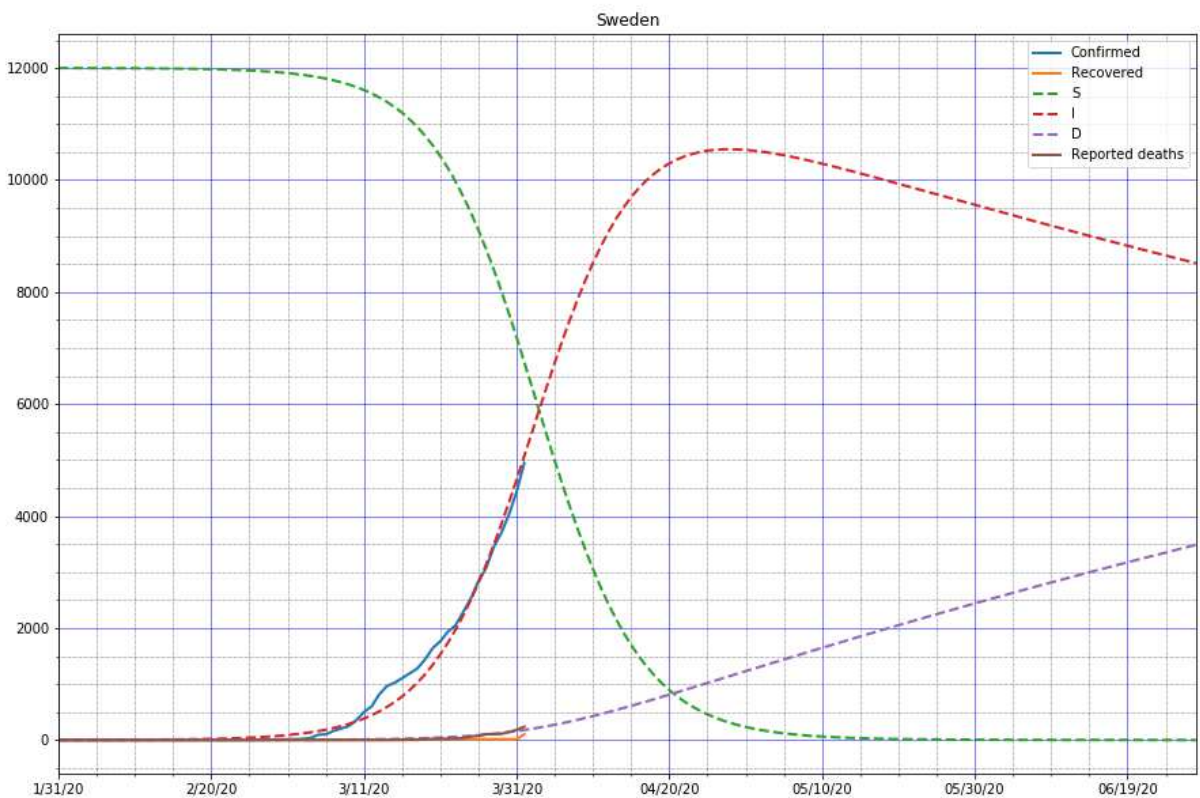}
		\caption{SIR model description of COVID-19 infected and recovery cases in Sweden as of April, 01. We take $N=12000$.
		}\label{sid}
	\end{center}
\end{figure*}

What we can see from above simulation is that the recovery rate will remain very low during the expected pandemic time interval. Instead we now include the reported death cases in above SIR model by replacing the quantity $R$ with $D$.  The result as of April 01 is given in Fig. \ref{sid}. The projected death cases are again very large and are more than 3000.

\section{Remarks as of April 02}
Our simulations show that all SI, SIR, SID models describe well the reported infected cases show rather modest increase in the near future which is very promising. 
However, the projected deceased cases in both SI and SID models are extremely high even though the uncertainty is astonishingly large. We deem it urgent to explore the uncertainty of our model. The new data from the future few days can be critical for confining the predicted curve on decreased cases. If the model is correct, one should worry that the infected cases in Sweden may be much much higher than it is reported today and a massive testing on exposed and suspicious cases may be urgent.

\end{document}